\documentclass{article}

\usepackage{hyperref}
\usepackage{alltt,color}
\usepackage{amsmath}
\usepackage[T1]{fontenc}
\usepackage{upquote}
\usepackage{listings}
\usepackage{lstcoq}
\definecolor{darkgreen}{rgb}{0.1,0.5,0.1}
\definecolor{darkred}{rgb}{0.7,0.1,0.1}
\definecolor{darkyellow}{rgb}{0.5,0.5,0.1}
\definecolor{darkblue}{rgb}{0.1,0.1,0.5}
\definecolor{darkgrey}{rgb}{0.125,0.125,0.125}

\definecolor{dkviolet}{rgb}{.5,0,.5}
\definecolor{dkblue}{rgb}{0,0,.5}
\definecolor{dkgreen}{rgb}{0,.5,0}
\definecolor{dkred}{rgb}{.5,0,0}
\definecolor{ltblue}{rgb}{0,.8,1}

\newcommand{\sNon}[1]{\ensuremath{\langle#1\rangle}}
\newcommand{\sTerm}[1]{\ensuremath{``#1"}}
\newcommand{\sStar}[1]{\ensuremath{\{ #1 \}}}
\newcommand{\sAlt}[2]{\ensuremath{ #1 \mid #2 }}
\newcommand{\sConc}[2]{\ensuremath{#1, #2}}
\newcommand{\sPar}[1]{\ensuremath{(#1)}}
\newcommand{\sProof}{\sNon{proof}}
\newcommand{\sLine}{\sNon{line}}
\newcommand{\sRat}{\sNon{rat}}
\newcommand{\sDelete}{\sNon{delete}}
\newcommand{\sId}{\sNon{id}}
\newcommand{\sRes}{\sNon{res}}
\newcommand{\sIdlist}{\sNon{idlist}}
\newcommand{\sClause}{\sNon{clause}}
\newcommand{\sPos}{\sNon{pos}}
\newcommand{\sNeg}{\sNon{neg}}
\newcommand{\sLit}{\sNon{lit}}

\author{Lu\'{i}s Cruz-Filipe \and Marijn Heule \and Warren Hunt \and Matt Kaufmann \and Peter Schneider-Kamp}
\title{Efficient Certified RAT Verification}

\begin{document}

\maketitle

\begin{abstract}
Clausal proofs have become a popular approach to validate the results
of SAT solvers. However, validating clausal proofs in the most widely
supported format (DRAT) is expensive even in highly optimized
implementations.  We present a new format, called LRAT, which extends
the DRAT format with hints that facilitate a simple and fast
validation algorithm.  Checking validity of LRAT proofs can be
implemented using trusted systems such as the languages supported by
theorem provers.  We demonstrate this by implementing two certified
LRAT checkers, one in Coq and one in ACL2.
\end{abstract}

\section{Introduction}

Consider a {\em formula}, or set of {\em clauses} implicitly
conjoined, where each clause is a list of {\em literals} (Boolean
proposition letters or their negations), implicitly disjoined.
Satisfiability (SAT) solvers decide the question of whether a given
formula is {\em satisfiable}, that is, true under some assignment of
{\em true} and {\em false} values to the Boolean proposition letters
of the formula.  SAT solvers are used in many applications in academia
and industry, for example to check the correctness of hardware and
software.  
A bug in such a SAT solver could result in an invalid claim that some
hardware or software model is correct.  In order to deal with this
trust issue, we believe a SAT solver should produce a proof of
unsatisfiability. In turn, this proof can and should be validated with
a trusted checker.

Early work on proofs of unsatisfiability focused on resolution
proofs. In short, a resolution proof states for each new clause how to
construct it via resolution steps. Resolution proofs are easy to
validate, but difficult to produce from today's SAT solvers. Moreover,
several state-of-the-art solvers use techniques that go beyond
resolution and therefore cannot be expressed using resolution proofs.

An alternative method is to produce {\em clausal proofs}, that is,
sequences of steps that each modify the current formula by specifying
the deletion of an existing clause or the addition of a new clause.
Such proofs are
supported by all state-of-the-art SAT solvers.  The most widely
supported clausal proof format is called DRAT, which is the format
required by the recent SAT competitions.  The DRAT proof format was
designed to make it as easy as possible to produce proofs, in order
to make it easy for implementations to support it.  DRAT checkers
increase the confidence in the correctness of unsatisfiability
results, but there is still room for improvement, i.e., by checking
the result using a highly-trusted system.

Our tool chain works as follows.  When a SAT solver produces a clausal proof
of unsatisfiability for a given formula, we validate this proof using
a fast non-certified proof checker, which then produces an optimized
proof with hints.  Then, using a certified checker, we validate that
the optimized proof is indeed a valid proof for the original formula.
We do not need to trust whether the original proof is correct.  In
fact, the non-certified checker might even produce an optimized proof
from an incorrect proof.

Validating clausal proofs is potentially expensive.  For each clause
addition step in a proof of unsatisfiability, unit clause propagation
(explained below) should
result in a conflict when performed on the current formula, based on
an assignment obtained by negating the clause to be added.
Thus, we may need to propagate thousands of unit clauses to check the
validity of a single clause addition step.  Scanning over the formula
thousands of times for a single check would be very expensive.  This
problem has been mitigated through the use of watch pointers.
However, validating clausal proofs is often costly even with watch pointers.

In this paper we first present the new expressive proof format LRAT
and afterwards show that this proof format enables the development of
efficient certified proof checkers.
This work builds upon previous work of some of the co-authors
\cite{grit}, as the LRAT format and the certified Coq checker
presented here extend the GRIT format and the certified Coq checker
presented there, respectively. Additionally, we implemented an efficient 
certified checker in the ACL2 theorem proving system.

The LRAT format poses several restrictions on the syntax in order to
make validation as fast as possible.  Each clause in the proof
must be suitably sorted.  This allows a simple check that
the clause does not contain duplicate or complementary
literals.  Hints are also sorted in such a way that they become unit
from left to right.  Finally, resolution candidates are sorted by
increasing clause index; this allows scanning the formula once.

This paper is structured as follows. In Section~\ref{sec:drat} we shortly recapitulate
the checking procedure for clausal proofs based on the DRAT format.
The novel LRAT format is introduced in Section~\ref{sec:lrat}.
We demonstrate the benefits of LRAT by extracting two certified checkers 
for the format: one in Coq (Section~\ref{sec:coq}) and one in ACL2 (Section~\ref{sec:acl2}).
We draw some conclusions in Section~\ref{sec:conclusion}.

\section{Background on Clausal Proof Checking}
\label{sec:drat}

Each step in a clausal proof is either the addition or the deletion of
a clause.  Each clause addition step should be {\em redundant},
that is, it should preserve satisfiability;
this should be checkable in polynomial time.  The polynomial time
checking procedure is described in detail below.  Clause deletion
steps need not be checked, because they trivially preserve
satisfiability.  The main reason to include clause deletion steps in
+ %
proofs is to reduce the computational and memory costs to validate
proofs.

A clause with only one literal is called a unit clause. Checking
whether a clause is redundant with respect to a CNF formula is
computed via Unit Clause Propagation (UCP). UCP works as follows:
For each unit clause $(l)$ all literal occurrences of $\bar l$ are removed
from the formula. Notice that this can result in new unit clauses.
UCP terminates when either no literals can be removed or when it
results in a conflict, i.e., all literals in a clause have been
removed.

Let $C$ be a clause. $\overline{C}$ denotes the negation of a clause, which
is a conjunction of all negated literals in $C$. A clause C has the
redundancy property Asymmetric Tautology (AT) with respect to a CNF
formula $F$ iff UCP on $F \land (\overline C)$ results in a conflict. The core
redundancy property used in the DRAT format is Resolution Asymmetric
Tautology (RAT). A clause $C$ has the RAT property with respect to a
CNF formula $F$ if there exists a literal $l \in C$ such that for all
clauses $D$ in $F$ with $\lnot l \in D$, the clause $C \lor (D \setminus \{\lnot l\})$
has the property AT with respect to $F$. Notice that RAT property is a
generalization of the AT property.

The DRAT proof checking works as follows. Let $F$ be the input formula
and $P$ be the clausal proof. At each step $i$, the formula is modified.
The initial state is: $F_{0} = F$. At step $i > 0$, the $i^{th}$ line of $P$ is
read. If the line has the prefix {\tt d}, then the clause C described on
that line is removed: $F_{i} = F_{i-1} \setminus \{C\}$. Otherwise, if there is
no prefix, then C must have the RAT property with respect to formula
$F_{i-1}$. This must be validated. Recall that the RAT property requires
a pivot literal $l$. In the DRAT formula it is expected that the first
literal in $C$ is the pivot. If the RAT property can be validated, then
the clause is added to the formula: $F_{i} = F_{i-1} \land C$. If the
validation fails, then the proof is invalid.

The empty clause, typically at the end of the proof, should have the AT property
as it does not have a first literal.

\section{Introducing the LRAT Format}
\label{sec:lrat}

The Linear RAT (LRAT) proof format is based on the
RAT property, and it is designed to make proof
checking as straightforward as possible.  The purpose of LRAT proofs
is to facilitate the implementation of proof validation software
using highly trusted systems such as theorem provers.
An LRAT proof can be produced when checking a DRAT proof
with a non-certified checker (cf.\ the end of this section).

The most costly operation during clausal proof validation is finding
the unit clauses during unit propagation.  The GRIT format~\cite{grit}
removes this problem by requiring proofs to include hints that list
all unit clauses.  This makes it much easier and faster to validate
proofs, because the checker no longer needs to find the unit clauses.
However, the GRIT format does not allow checking of all possible clauses that
can be learned by today's SAT solvers and expressible in the DRAT format.  The LRAT format extends the
GRIT format to remove this limitation.

Unlike the GRIT format, the LRAT format supports checking clauses with
the RAT property.  To check such a
clause, a {\em pivot element} is chosen from it, and
then the RAT property is checked for all clauses containing the
negation of the pivot element.  In order to enable efficient RAT checking
the LRAT format requires that all clauses containing the negated-pivot
element be specified.  Furthermore, for each resolvent it has to be
specified how to perform UCP as is done for AT in the GRIT approach.

While the LRAT format is semantically an extension of the GRIT format,
we updated two aspects.  First, the clauses from the original CNF are
\emph{not} included, as this required verification that these clauses
do indeed occur in the original CNF.  The advantage of working only
with a subset of clauses from the original CNF can be achieved by
starting with a deletion step for clauses not
relevant for the proof.  Second, the syntax of the deletion
information has been extended to include a clause identifier.  To be recognized,
deletion statements are now identified with lines that start with
an index followed by ``d''.
This change makes the format stable
under permutations of lines.  In practice, checkers expect proof
statements in ascending order, which easily can be achieved by sorting
numerically, e.g., using ``\texttt{sort -n}''.

To demonstrate these two changes, we first consider an example, which
does \emph{not} use the RAT property.  \autoref{fig:norat} shows an
original CNF, the DRUP proof obtained by a SAT solver, the GRIT
version of that proof, and, finally, the equivalent LRAT proof.

\begin{figure}[t]
\small
  \begin{minipage}[t]{0.175\textwidth}
CNF formula%
    \begin{alltt}
p cnf 4 8
\textcolor{darkgreen}{ 1  2 -3 0}
\textcolor{darkgreen}{-1 -2  3 0}
\textcolor{darkgreen}{ 2  3 -4 0}
\textcolor{darkgreen}{-2 -3  4 0}
\textcolor{darkgreen}{-1 -3 -4 0}
\textcolor{darkgreen}{ 1  3  4 0}
\textcolor{darkgreen}{-1  2  4 0}
\textcolor{darkgreen}{ 1 -2 -4 0}
    \end{alltt}
  \end{minipage}
  \begin{minipage}[t]{0.195\textwidth}
DRUP format%
    \begin{alltt}
  \textcolor{darkred}{    1  2 0}
d \textcolor{darkblue}{ 1 -3  2} 0
  \textcolor{darkred}{    1  3 0}
d \textcolor{darkblue}{ 1  4  3} 0
  \textcolor{darkred}{       1 0}
d \textcolor{darkblue}{    1  3} 0
d \textcolor{darkblue}{    1  2} 0
d \textcolor{darkblue}{ 1 -4 -2} 0
  \textcolor{darkred}{       2 0}
d \textcolor{darkblue}{-1  4  2} 0
d \textcolor{darkblue}{ 2 -4  3} 0
  \textcolor{darkred}{         0}
    \end{alltt}
  \end{minipage}
  \begin{minipage}[t]{0.32\textwidth}
GRIT format%
    \begin{alltt}
 1 \textcolor{darkgreen}{ 1  2 -3 0} 0
 2 \textcolor{darkgreen}{-1 -2  3 0} 0
 3 \textcolor{darkgreen}{ 2  3 -4 0} 0
 4 \textcolor{darkgreen}{-2 -3  4 0} 0
 5 \textcolor{darkgreen}{-1 -3 -4 0} 0
 6 \textcolor{darkgreen}{ 1  3  4 0} 0
 7 \textcolor{darkgreen}{-1  2  4 0} 0
 8 \textcolor{darkgreen}{ 1 -2 -4 0} 0
 9 \textcolor{darkred}{1 2 0}       \textcolor{darkyellow}{1 6 3} 0
                 0 \textcolor{darkblue}{1} 0
10 \textcolor{darkred}{1 3 0}       \textcolor{darkyellow}{9 8 6} 0
                 0 \textcolor{darkblue}{6} 0
11   \textcolor{darkred}{1 0}    \textcolor{darkyellow}{10 9 4 8} 0
            0 \textcolor{darkblue}{10 9 8} 0
12   \textcolor{darkred}{2 0}    \textcolor{darkyellow}{11 7 5 3} 0
               0 \textcolor{darkblue}{7 3} 0
13     \textcolor{darkred}{0} \textcolor{darkyellow}{11 12 2 4 5} 0
    \end{alltt}
  \end{minipage}
  \begin{minipage}[t]{0.2425\textwidth}
LRAT format%
    \begin{alltt}
 9 \textcolor{darkred}{1 2 0}       \textcolor{darkyellow}{1 6 3} 0
 9               d \textcolor{darkblue}{1} 0
10 \textcolor{darkred}{1 3 0}       \textcolor{darkyellow}{9 8 6} 0
10               d \textcolor{darkblue}{6} 0
11   \textcolor{darkred}{1 0}    \textcolor{darkyellow}{10 9 4 8} 0
11          d \textcolor{darkblue}{10 9 8} 0
12   \textcolor{darkred}{2 0}    \textcolor{darkyellow}{11 7 5 3} 0
12             d \textcolor{darkblue}{7 3} 0
13     \textcolor{darkred}{0} \textcolor{darkyellow}{11 12 2 4 5} 0
    \end{alltt}
  \end{minipage}
  \caption{A CNF formula and three similar proofs of unsatisfiability
    in the DRUP, GRIT and LRAT format, respectively.  Formula clauses
    are shown in green, deletion information in blue, learned clauses
    in red, and unit propagation information in yellow.  The proofs do
    not have clauses based on the RAT property.  The spacing shown aims to
    improve readability, but extra spacing does not effect the meaning
    of a LRAT file.}
  \label{fig:norat}
\end{figure}

To specify a redundant clause with the RAT property, we extend the
format used for the AT property in GRIT.  The line starts with the
clause identifier of the new clause followed by the 0-terminated new
clause.  The first literal of the new clause is required to be the
pivot literal. Next, for each clause with clause identifier $i$
containing the negated-pivot element, we specify the (negative)
integer $-i$ followed by a (possibly empty) list of (positive) clause
identifiers used in UCP of the new clause with clause $i$.

\begin{figure}[t]
  \begin{minipage}[t]{0.205\textwidth}
CNF formula%
    \begin{alltt}
p cnf 4 8
\textcolor{darkgreen}{ 1  2 -3 0}
\textcolor{darkgreen}{-1 -2  3 0}
\textcolor{darkgreen}{ 2  3 -4 0}
\textcolor{darkgreen}{-2 -3  4 0}
\textcolor{darkgreen}{-1 -3 -4 0}
\textcolor{darkgreen}{ 1  3  4 0}
\textcolor{darkgreen}{-1  2  4 0}
\textcolor{darkgreen}{ 1 -2 -4 0}
    \end{alltt}
  \end{minipage}
~~~
  \begin{minipage}[t]{0.255\textwidth}
 DRAT format%
    \begin{alltt}
   \textcolor{darkred}{      1 0}
d \textcolor{darkblue}{ 1 -4 -2 0}
d \textcolor{darkblue}{ 1  4  3 0}
d \textcolor{darkblue}{ 1  2 -3 0}
   \textcolor{darkred}{      2 0}
d \textcolor{darkblue}{-1  2  4 0}
d \textcolor{darkblue}{ 2 -4  3 0}
   \textcolor{darkred}{        0}
    \end{alltt}
  \end{minipage}
~~~
  \begin{minipage}[t]{0.2425\textwidth}
LRAT format%
    \begin{alltt}
 9 \textcolor{darkred}{1 0} \textcolor{cyan}{-2} \textcolor{darkyellow}{6 8} \textcolor{cyan}{-5} \textcolor{darkyellow}{1 8} \textcolor{cyan}{-7} \textcolor{darkyellow}{6 1} 0
 9                  d \textcolor{darkblue}{8 6 1} 0
10 \textcolor{darkred}{2 0}              \textcolor{darkyellow}{9 7 5 3} 0
10                    d \textcolor{darkblue}{7 3} 0
11 \textcolor{darkred}{  0}           \textcolor{darkyellow}{9 10 2 4 5} 0
    \end{alltt}
  \end{minipage}
  \caption{The LRAT format with the RAT property (with original clauses in green, deletion information in blue, learned clauses in red, unit propagation information in yellow, and resolution clauses in cyan).}
  \label{fig:rat}
\end{figure}

For example, consider the first line of the LRAT proof in Figure~\ref{fig:rat}:
\begin{alltt}
\begin{center}
9 \textcolor{darkred}{1 0} \textcolor{cyan}{-2} \textcolor{darkyellow}{6 8} \textcolor{cyan}{-5} \textcolor{darkyellow}{1 8} \textcolor{cyan}{-7} \textcolor{darkyellow}{6 1} 0
\end{center}
\end{alltt}
The first number, {\tt 9} expresses that the new clause will get
identifier 9. The numbers in between the identifier and the first {\tt
  0} express the literals in the clause. In clause of clause {\tt 9}
this is only literal {\tt 1}. After the first {\tt 0} follow the
hints. All hints are clause identifiers. Positive hints express that
the clause becomes unit or falsified.  Negative hints express that the
clause is a candidate for a RAT check, i.e., it contains
the complement of the pivot element. In the example line, there are
three such negative hints: {\tt \textcolor{cyan}{-2}}, {\tt
  \textcolor{cyan}{-5}}, and {\tt \textcolor{cyan}{-7}}. The LRAT
format prescribes that negative literals are listed in increasing
order of their absolute value.

After a negative hint there may be positive hints that list the
identifiers of clauses that become unit and eventually falsified. For
example, assigning the literal in the new clause ({\tt 1}) to false as
well as the literals in the second clause apart from the pivot ({\tt
  2} and {\tt -3}), then clause six becomes unit ({\tt 4}), which in
turn falsifies clause eight.

There are two extensions to this kind of simple RAT checking.
(1) It is possible that there are no positive hints following a negative
hint. In this case, the new clause and the candidate for a RAT check have
two pairs of complementary literals. (2) It is also possible that some
positive hints are listed before the first negative hint. In this
case, these clauses (i.e., whose identifiers are listed) become
unit after assigning the literals in the new clause to false.

The full syntax of the LRAT format is given by the grammar in
\autoref{fig:syntax}, where for the sake of sanity, whitespace (tabs
and spaces) is ignored.  Note that syntactically, AT and RAT lines are
both covered by RAT lines. AT is just the special case where there is
a non-empty list of only positive hints.
\begin{figure}[t]
\begin{center}
\begin{tabular}{lcl}
\sProof & = & \sStar{\sLine}\\
\sLine & = & \sConc{\sPar{\sAlt{\sRat}{\sDelete}}}{\sTerm{\backslash n}}\\
\sRat & = & \sConc{\sId}{\sConc{\sClause}{\sConc{\sTerm{0}}{\sConc{\sConc{\sIdlist}{\sStar{\sRes}}}{\sTerm{0}}}}}\\
\sDelete & = & \sConc{\sId}{\sConc{\sTerm{d}}{\sConc{\sIdlist}{\sTerm{0}}}}\\
\sRes & = & \sConc{\sNeg}{\sIdlist}\\
\sIdlist & = & \sStar{\sId}\\
\sId & = & \sPos\\
\sLit & = & \sAlt{\sPos}{\sNeg}\\
\sPos & = & \sAlt{\sTerm{1}}{\sAlt{\sTerm{2}}{\ldots}}\\
\sNeg & = & \sConc{\sTerm{\text{-}}}{\sPos}\\
\sClause & = & \sConc{\sStar{\sLit}}{\sTerm{0}}\\
\end{tabular}
\end{center}
  \caption{EBNF grammar for the LRAT format.}
  \label{fig:syntax}
\end{figure}

Producing LRAT proofs directly from SAT solvers would add significant
overhead both in runtime and memory usage, and it might require the
addition of complicated code.  Instead, we extended the DRAT-trim
proof checker \cite{drat-trim} to emit LRAT proofs.  DRAT-trim already
supported the emitting of optimized proofs in the DRAT and TraceCheck+
formats.  DRAT-trim emits an LRAT proof after validation of a proof
using the ``{\tt -L proof.lrat}'' option.

We implemented an uncertified checker for LRAT in C that achieves
runtimes comparable to the one from \cite{grit} on examples without
RAT lines.

\section{Extending the GRIT Checker to LRAT}
\label{sec:coq}

In this section we extend the formalization of the GRIT checker from \cite{grit} to the whole syntax of LRAT by adding results about the RAT property. We assume familiarity with \cite{grit}.
Due to the need to consider extension (1) discussed in the previous section and its combination with extension (2), these results are a bit more
complicated than the ones previously needed.

\begin{lstlisting}
Lemma RAT_lemma_1 : forall (c:CNF) (l:Literal) (cl:Clause),
  (forall (cl':Clause), CNF_in cl' c -> 
      (entails c ((remove literal_eq_dec (negate l) cl') ++ cl))
      \/ (exists l', l'<>l /\ In (negate l') cl' /\ (In l' cl \/ entails c (negate l'::l::cl))))
  -> forall V, satisfies V c -> exists V, satisfies V (CNF_add (l::cl) c).
\end{lstlisting}

In this lemma, \lstinline+c+ is the CNF we start with, and \lstinline+l::cl+ is the clause for which we want to verify
the RAT property with respect to \lstinline+c+.
(We single out the pivot \lstinline+l+.)
The hypothesis states that, for every clause \lstinline+cl'+ in \lstinline+c+, either \lstinline+c+ entails the clause
obtained by removing $\neg$\lstinline+l+ from \lstinline+cl'+ and joining with \lstinline+cl+, or there exists a
literal \lstinline+l'+, distinct from the pivot, whose negation is in \lstinline+cl'+, and such that either \lstinline+l'+
occurs in \lstinline+cl+ or \lstinline+c+ entails the disjunction of $\neg$\lstinline+l'+ and \lstinline+l::cl+.

Observe that the quantification is over \emph{all} the formulas in \lstinline+c+, rather than over those containing
$\neg$\lstinline+l'+ (as required by the RAT property): for formulas not containing $\neg$\lstinline+l'+ the first case
trivially holds, and this formulation is simpler.

\begin{lstlisting}
Lemma RAT_lemma_2: forall l c cl cl', CNF_in cl' c -> ~(In (negate l) cl') ->
      entails c ((remove literal_eq_dec (negate l) cl') ++ cl).
\end{lstlisting}

We then define our iterative function performing the RAT check.
We refer to~\cite{grit} for the discussion of the different representations for clauses and CNFs.
The argument to \lstinline+RAT_check+ has type \lstinline+ICNF+, which implements a CNF as a \lstinline+Map+ (identified
by an index, as in the GRIT format).
It is transformed in a list, over which we do iteration in the auxiliary function \lstinline+RAT_check_run+.
Finally, the list \lstinline+L+ provides the witnesses for each RAT check.
It is a list of pairs having a clause identifier as first argument and either a list of clauses (used for unit
propagation to establish the first possible valid case of the RAT check) or a literal (the duplicate literal in the
second case) together with a list of clauses used again for unit propagation to establish the second case.
For legibility, we omit several proof terms in the code below.

\begin{lstlisting}
Definition RAT_check (c:ICNF) (pivot:Literal) (cl:Clause)
                     (L:list (N*((list N)+(Literal*(list N))))) :=
  RAT_check_run c (ICNF_to_list c) pivot (Clause_to_SetClause cl) L.

Fixpoint RAT_check_run (c:ICNF) (c':list (N*{cl:SetClause | SC_wf cl}))
         (pivot:Literal) (cl:SetClause) (L:list (N*((list N)+(Literal*(list N))))) :=
  match c' with
  | nil => true
  | (i,(exist cl' Hcl'))::newC =>
       if (BT_in_dec _ _ _ _ (negate pivot) cl' Hcl')
       then let LIST := get_list_from i L in
            match LIST with
            | inl is => (propagate c ((BT_add_all _ _ (BT_remove _ (negate pivot) cl')
                                                       (BT_add _ pivot cl))) is)
                        && (RAT_check_run c newC pivot cl L)
            | inr (lit,is) => match literal_eq_dec pivot lit with
                  | left _ => false
                  | right _ => (SC_has_literal (negate lit) cl' Hcl')
                               && (C_has_literal lit (SetClause_to_Clause cl)
                                   || propagate c (BT_add _ (negate lit) (BT_add _ pivot cl)) is)
                               && (RAT_check_run c newC pivot cl L)
            end end
       else RAT_check_run c newC pivot cl L
  end.
\end{lstlisting}

The proof is technical, and for convenience divided into several lemmas.
The main theorem states that, if the RAT check succeeds, then we can add the required clause to the CNF preserving
satisfiability.

\begin{lstlisting}
Theorem RAT_theorem : forall c pivot cl L, RAT_check c pivot cl L = true -> 
        forall V, satisfies V c ->
        exists V, satisfies V (CNF_add (Clause_to_SetClause (pivot::cl)) (ICNF_to_CNF c)).
\end{lstlisting}

In order to use this result and check proofs of unsatisfiability that use the RAT property, we enrich the type of
actions provided by the oracle.\footnote{We also changed the algorithm slightly from~\cite{grit}: the working set is now
initialized to contain the original CNF, which allows us to remove the action ``add a formula from the original CNF to
the working set''.}

\begin{lstlisting}
Inductive Action : Type :=
  | D : list ad -> Action
  | R : ad -> Clause -> list ad -> Action
  | A : ad -> Literal -> Clause -> list (ad * ((list ad)+(Literal*(list ad)))) -> Action.
\end{lstlisting}

In the definition of \lstinline+refute_work+, we add the corresponding case for the new type of action.

\begin{lstlisting}
Function refute_work (w:ICNF) (O:Oracle)
  {measure Oracle_size O} : Answer :=
  match (force O) with
  ...
  | lcons (A i p cl L) O' => andb (RAT_check w p cl L)
                                  (refute_work (add_ICNF i (p::cl) _) w) O')
  end.
\end{lstlisting}

The proof of soundness simply requires checking the extra case, and we obtain the same results as before.

\begin{lstlisting}
Lemma refute_work_correct : forall w O, refute_work w O = true -> unsat w.

Definition refute (c:list (ad * Clause)) (O:Oracle) : Answer :=
  refute_work (make_ICNF c) O.

Theorem refute_correct : forall c O, refute c O = true -> unsat (make_ICNF c).
\end{lstlisting}

By extracting \lstinline+refute+ we again obtain a correct-by-construction checker for proofs of unsatisfiability using
the full LRAT format.
If this checker returns \lstinline+true+ when given a particular CNF and proof, this guarantees that the CNF is indeed
unsatisfiable.
The universal quantification over the oracle ensures that any errors in its implementation (and in particular in the
interface connecting it to the checker) do not affect the correctness of this answer.

\paragraph{Entailment checking.}
If the size of a proof is enormous, proof checking will be expensive even for LRAT proofs. In order to make proof 
checking feasible in reasonable time, one can check the proof in parallel. This can be achieved by partitioning a proof 
and verifying each part independently~\cite{cpp}. Let $P$ be a proof of unsatisfiability for a CNF formula $F_0$. 
We can partition $P$ into $k$ parts $\{P_1, \dots, P_k\}$. The formulas $F_i$ with $i \in \{1, \dots, k\}$ are defined as
applying (but not verifying) proof $P_i$ to formula $F_{i-1}$. In order to verify that $P$ is a valid proof for $F_0$, 
it is sufficient to show that all steps in the proof $P_i$ are valid for formula $F_{i-1}$ and that formula $F_i$ is
entailed by the formula obtained by applying $P_i$ to $F_{i-1}$. Finally, one of $P_1, \dots, P_k$ should contain the empty clause.

To validate a partial proof, we want to verify a reduction. %
In other words, starting from a CNF, we apply and verify a sequence of actions described in the LRAT format. %
In this case, we know that satisfiability of the starting CNF ($F_{i-1}$) implies satisfiability of the resulting CNF ($F_i$).

In order to deal with partial proof checking, we tweak our definition of \lstinline+refute_work+ slightly to return a
pair consisting of a boolean value and a CNF.
The base case is changed: when there are no more actions, we return \lstinline+true+ (instead of \lstinline+false+)
together with the CNF currently stored.
When we derive the empty clause, we also return \lstinline+true+, but this time together with a CNF containing only the
empty clause.
In the remaining cases, if any test fails we return \lstinline+false+ with the formula currently stored; otherwise we
propagate the result from the recursive call.

The soundness of the main function now looks as follows.

\begin{lstlisting}
Definition ICNF_reduces (C C':ICNF) := forall V, satisfies V (ICNF_to_CNF C) ->
  exists V', satisfies V' (ICNF_to_CNF C').

Lemma refute_work_correct : forall w O F, refute_work w O = (true,F) -> ICNF_reduces w F.
\end{lstlisting}

Since we can test whether a formula is a CNF containing only the empty clause, we can immediately derive the original
implementation of \lstinline+refute+ and reprove its soundness.

\begin{lstlisting}
Definition refute (c:list (ad * Clause)) (O:Oracle) : bool :=
  let (b,F) := refute_work (make_ICNF c) O in
  b && (if (ICNF_eq_empty_dec F) then true else false).

Theorem refute_correct : forall c O, refute c O = true -> unsat (make_ICNF c).
\end{lstlisting}

Furthermore, we can provide a target CNF and check that the oracle provides a correct reduction from the initial CNF to
the target.

\begin{lstlisting}
Definition entail (c c':list (ad * Clause)) (O:Oracle) : bool :=
  let (b,F) := refute_work (make_ICNF c) O in
  b && (if (ICNF_all_in_dec (map snd c') _ (ICNF_to_CNF_wf F)) then true else false).

Theorem entails_correct : forall c c' O, entail c c' O = true ->
  ICNF_reduces (make_ICNF c) (make_ICNF c').
\end{lstlisting}

\paragraph{Results.}
After adapting the interface to be able to transform proofs in the full LRAT format into the oracle syntax defined
above, we tested the extracted checker on several unsatisfiability proofs output by SAT solvers supporting that format.

We also used the possibility of verifying entailments to check the transformation proof from~\cite{heule-sat16a}, the
only SAT-related step in the original proof of the Boolean Pythagorean Triples problem that we were unable to verify
in~\cite{grit}. The certified LRAT checker in Coq was able to verify this proof in $8$ minutes and $25$ seconds, including
approx.\ $15$ seconds for the entailment checking.

\section{LRAT Checker in ACL2}
\label{sec:acl2}

In this section, in order to demonstrate the general applicability of our approach,
we extended the ACL2-based DRAT checker from \cite{nathan} to permit the
checking of UNSAT proofs in the LRAT format. We have certified this extension
using the ACL2 theorem-proving system. 
We outline our
formalization below using the Lisp-style ACL2 syntax, with comments to
assist readers unfamiliar with Lisp syntax.  Note that embedded
comments begin with a ``{\bf ;}'' character and continue to the end of
a line.

We omit the code here but note that it has been optimized for
efficiency, in particular using applicative hash tables for formulas
that are heuristically cleaned on occasion after deletion.  Of course,
correctness of such optimizations was necessarily proved as part of
the overall correctness proof.  The code and top-level theorem are
available
from the top-level file {\tt top.lisp} in the full proof
development~\cite{acl2-lrat}, included in the GitHub
repository~\cite{acl2-github} that holds ACL2 and its libraries.  Also
see the {\tt README} file in that directory.  Here
we focus primarily on the statement of correctness.

The top-level correctness theorem is as follows.
\begin{verbatim}
(defthm main-theorem
  (implies (and (formula-p formula)             ; Valid formula and
                (refutation-p proof formula))   ; Valid proof with empty clause
           (not (satisfiable formula))))        ; Imply unsatisfiable
\end{verbatim}
The command {\tt defthm} is an ACL2 system command that demands that
the ACL2 theorem-proving system establish the validity of the claim
that follows the name (in this case {\tt main-theorem}) of the theorem
to be checked.

The theorem above is expressed in terms of functions {\tt formula-p},
{\tt refutation-p}, and {\tt satisfiable}.  The first of these
recognizes structures that represent sets of clauses; our particular
representation uses applicative hash tables~\cite{fast-alists}.
The function {\tt refutation-p} recognizes valid proofs that yield a
contradiction; thus, it calls other functions, including one that
performs the necessary RAT checks.  We verify a proof by checking that
each step of an alleged proof redundantly extends a given formula.

Finally, we define {\tt satisfiable} to mean that there exists an
assignment satisfying a given formula.  The first definition says that
the given assignment satisfies the given formula, while the second
uses an existential quantifier to say that {\em some} assignment
satisfies the given formula.
\begin{verbatim}
(defun solution-p (assignment formula)
  (and (clause-or-assignment-p assignment)
       (formula-truep formula assignment)))

(defun-sk satisfiable (formula)
  (exists assignment (solution-p assignment formula)))
\end{verbatim}

Before our SAT proof checker can be called, an LRAT-style proof is
read from a file, and during the reading process it is converted into
an internal Lisp format that is used by our checker.  Using the ACL2
theorem prover, we have verified the theorem {\tt main-theorem} above,
which states that our code correctly checks the validity of a proof of
the empty clause.

\paragraph{Results.}
The ACL2 checker %
is able to check the validity of adding each of the $68{,}667$ clauses in the transformation
proof from \cite{heule-sat16a}
in less than $9$ seconds. The certified checking of this LRAT proof is almost as
fast as non-certified checking and conversion of the DRAT proof into the LRAT proof
by DRAT-trim. This is a testament to the efficiency potential of the LRAT
format in particular, and the approach taken in our work in general. At the moment 
of writing, the entailment checking has not been implemented yet to the ACL2 
checker, but this can easily be added in a similar way as we did for the Coq checker.

\section{Conclusions}
\label{sec:conclusion}

We have introduced a novel format for clausal proof checking, \emph{Linear RAT} (LRAT), 
which extends the GRIT format~\cite{grit} to support checking all techniques used in state-of-the-art SAT solvers.
We have shown that it allows for implementing efficient certified proof checkers for UNSAT proofs with the RAT property,
both using Coq and using ACL2. The ACL2 LRAT checker is almost as fast as ---and in some cases even faster than---
non-certified checking by DRAT-trim of the corresponding DRAT proof. This suggests that certified checking can
be achieved with a reasonable overhead. 

Furthermore, we have shown that our Coq checker's ability to check entailment and thereby transformation proofs has allowed us
to check the transformation proof from~\cite{heule-sat16a}, the
only SAT-related step in the original proof of the Boolean Pythagorean Triples problem that we were unable to verify
in~\cite{grit}.

\newpage

\bibliographystyle{abbrv}
\bibliography{paper}
\end{document}